\documentclass[aps,pre,showpacs,preprint,groupedaddress,amsmath,amssymb]{revtex4}

\usepackage{graphicx}

\begin{document}

\preprint{APS/XVS}
\title{Synchronization in Complex Systems Following the Decision Based Queuing Process: The Rhythmic Applause as a Test Case}


\author{D. Xenides}
\email{xenides@uop.gr}
\author{D. S. Vlachos}
\email{dvlachos@uop.gr}
\author{T. E. Simos}
\affiliation{Laboratory of Computational Sciences, Department of Computer Science and Technology, University of Peloponnese, GR-22 100, Terma Karaiskaki, Tripolis, Greece}



\date{\today}

\begin{abstract}
Living communities can be considered as complex systems, thus a
fertile ground for studies related to their statistics and dynamics.
In this study we revisit the case of the rhythmic applause by
utilizing the model proposed by V\'azquez et al. [A. V\'azquez et
al., Phys. Rev. E 73, 036127 (2006)] augmented with two contradicted
{\it driving forces}, namely: {\it Individuality} and {\it
Companionship}. To that extend, after performing computer
simulations with a large number of oscillators we propose an
explanation on the following open questions (a) why synchronization
occurs suddenly, and b) why synchronization is observed when the
clapping period ($T_c$) is $1.5 \cdot T_s < T_c < 2.0 \cdot T_s$
($T_s$ is the mean self period of the spectators) and is lost after
a time. Moreover, based on the model, a weak preferential attachment
principle is proposed which can produce complex networks obeying
power law in the distribution of number edges per node with exponent
greater than 3.
\end{abstract}




\maketitle

\section{Introduction}

Periodic phenomena are of high abundance in nature, what makes them
attractive to physicists is the fact that they can serve as
prototype complex networks, thus the importance on understanding
their statistics and dynamics. To give an idea about the
universality of the phenomenon we mention, among other systems, the:
{\it Pteroptyx malaccae} fireflies \cite{buck_science_159_1319_62},
secretory cells \cite{secretory_cell}, synchronous firing neurons
\cite{tsubo_prl_99_228101_07}, and rhythmic applause
\cite{barabasi_nature_403_849_00}. In particular, the study of the
mechanism that leads to self organization of biological systems
through synchronization is of major importance, since it provides
enough information as to understand the dynamics of the interactions
among the members of the system.

To address the problem of synchronization we used the coupled
oscillators approach. It is believed that Huygens ideas about the
synchronous move of wall hanged pendulums initiated the studies
towards the explanation of this phenomenon. It is noted that
coupling between identical oscillators is trivial since it can be
achieved by a phase-minimization procedure. This is not the case in
biological systems since the oscillators can be considered anything
but equivalent. In such diverse systems depending on the strength
and type of interaction as well as the dispersion of the self
periods, synchronization might or might not appear. To illustrate
the problem we refer to the synchronous clapping of an audience
after a spectacle. The audience is comprised by a large number of
spectators each one characterized by her/his temper, enthusiasm,
acoustic behavior, thus can be considered to applause in her/his
unique self period.

A first approach to the rhythmic applause indicates  that the
integrate-and-fire-type model
\cite{mirollo_siam_50_1645_90,bottani_pre_54_2334_97} is the most
promising to understand the synchronization mechanism. However, this
model doesn't take into account memory effects which are crucial for
the synchronization to occur, thus the continuous phase coupling in
the Kuramoto model \cite{kuramoto_JStatPhys_49_569_1987} has been
used to understand the underlying dynamics
\cite{barabasi_pre_61_6987_00}. On the other hand, a recent study
brought fourth a strong evidence  that human dynamics of many
social, technological and economic phenomena can be modeled by a
decision based queuing process \cite{barabasi_pre_73_036127_06} in
which individuals execute a task from a list of pending tasks based
on a given priority.

The purpose of the present paper is to show that application of the
aforementioned decision based queuing process is capable of modeling
the dynamics of the rhythmic applause and especially: i) why
synchronization occurs suddenly, ii) why synchronization is achieved
in a period almost two times the average self period of the
spectators, and iii) why synchronization appears and disappears
several times during the applause.

\section{Rhythmic applause and the Kuramoto model}

The Kuramoto model has been used to model the physics of the
rhythmic applause. In this model we deal with N--oscillators each of
them described by its $\phi_j$ phase. The oscillators have a
$g(\omega)$ distribution of their $\omega_j$ self frequency (where
$g(\omega)$ is considered to be a normal distribution). Every
rotator interacts with all other ones via
phase-difference-minimizing terms via the following formula
\begin{equation}
W_j^{int}=\frac{K}{N}\sum_{i=1}^N sin(\phi_i-\phi_j)
\end{equation}
The N coupled differential equations describing the over-damped
oscillator dynamics are
\begin{equation}
 \label{kuramoto}
\frac{d\phi_j}{dt} = \omega_j + \frac{K}{N} \sum_{i=1}^{N} sin(\phi_i-\phi_j)
\end{equation}
Mathematically the synchronization level will be characterized by an
order parameter, q, defined at any time moment as
\begin{equation}
q= \left | \frac{1}{N} \sum_{j=1}^Ne^{i\phi_j} \right |
\end{equation}
The maximal possible value q=1 corresponds to total synchronization,
the case $0<q<1$ to partial synchronization, while for q=0 there is
no synchronization at all in the system.

In the $N\rightarrow\infty$ thermodynamic limit of the equilibrium
dynamics ($t\rightarrow\infty$, so initial transient effects are
lost) Kuramoto and Nishikawa proved the existence of a $K_c$
critical coupling. For a Gaussian distribution of the oscillators'
natural frequencies, characterized by a dispersion $D$, they got

\begin{equation}
K_c = \sqrt{\frac{2}{\pi^3}D}
\end{equation}

For $K \leq K_c$, the only possible solution gives q=0 (no
synchronization) while for $K>K_c$ a stable solution with $q \neq 0$
appears. Thus, the main result is that for a population of globally
coupled nonidentical oscillators a partial synchronization of the
phases is possible whenever the interaction among oscillators
exceeds a critical value.

There are for four major drawbacks of this model when applied to
rhythmic applause:
\begin{enumerate}
\item The uniform one to one coupling between all the
oscillators. The spatial distribution of spectators prohibits this
assumption.

\item The coupling constant should be $K \ge K_c$. Since i) synchronization is achieved at a period $T_c
\approx 2\cdot T_s$ (with $T_s$ the mean self period of the
spectators), and ii) the doubling of the period leads to the
half-doubling of the dispersion coefficient ($D$), the following
inequality
\begin{equation}
\label{halfdoubling}
  1/2 \sqrt{2/\pi^3} D < K < \sqrt{2/\pi^3} D
\end{equation}
should hold. We note that $D$ is strongly related to behavioral
characteristics of the spectator (temperament, enthusiasm), thus
unrelated to $K$ which, in turns, is related to factors such as the
distribution of the sound in the room, reflections of the sound etc.
On the other hand the ``generalized relaxation oscillators``
approach is characterized, again, by the need of a global coupling
among the oscillators that can not be guaranteed in networks of
living organisms.

\item Usually at the beginning of the applause there is a long
"waiting" time without any synchronization, and with no increase in
the order parameter. Partial synchronization evolves suddenly after
that and achieves its maximal value in a short time. This should not
be the case if we are in the $K>K_c$ limit. One would expect in this
limit a continuous increase in the order parameter right from  the
beginning of the applause.

\item why is synchronization already achieved lost after a time, and
why might it reappear again? Loss of  synchronization should not
happen in the $K>K_c$ limit.
\end{enumerate}

\section{The decision based queuing process}

Recently, V\'azquez et al. \cite{barabasi_pre_73_036127_06} proposed
a decision based queuing process (DBQP) able to reproduce the
distribution of the possible action of humans in particular cases.
In this model, all the possible actions are forming a list and the
acted one is chosen by applying criteria (e.g., priority) related to
the specific individual, that is the emotional factors are playing
an equally important role.

Applying the above DBQP model in the case of the rhythmic applause,
the spectator she/he has to chose among the following actions
(tasks):
\begin{enumerate}
 \item to clap at the frequency that fulfill her/his satisfaction (this an expression of {\it Individualism}), or
 \item to try to change her/his pace  as to follow the clapping frequency of the synchronized spectators
 (this an expression of {\it Companionship}). It is obvious that this holds true only in the presence of such a
 group of people.
\end{enumerate}

\begin{figure}
\includegraphics[scale=0.5]{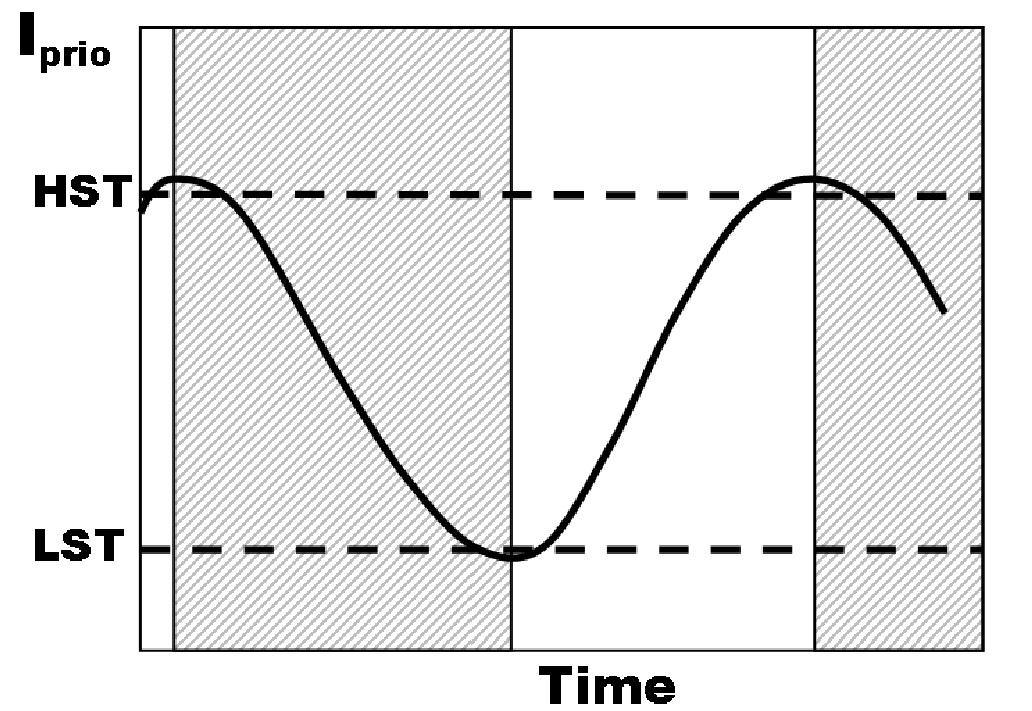}%
 \caption{\label{fig1:Indiv_Partn}}
 \end{figure}

Every time that a spectator claps her/his hands she/he makes a
decision between (1) and (2), depending on the priority given to
each of them. The priority is immediately related to the reached
level of satisfaction reflected to either the need of expressing
her/his {\it Individualism} or her/his {\it Companionship}. We name
the first as $I_{prio}$. The fluctuation of this parameter with the
time is ruled by the level of its saturation: when it is saturated
then $I_{prio}$ reduces with time, whereas increases in the other
case. Two boundaries control this behavior, namely the High
Saturation Threshold ($HST$) and the Low Saturation Threshold
($LST$). When $I_{prio}$ climbs over $HST$, then the spectator's
need to express her/his {\it Individualism} is considered to
saturate, thus the decay of $I_{prio}$. On the other hand when
$I_{prio}$ fall under $LST$, then the need to express the {\it
Individualism} comes again in the foreground, thus the $I_{prio}$
starts to increase. This mechanism is depicted in  Fig.
\ref{fig1:Indiv_Partn}.


A model system of a square grid with  $n \times n = n^2$ spectators
is chosen and it is assumed that the sound produced in position,
$(i,j)$ is heard weakened, proportional to the distance from the
source, in neighboring regions. Thus, the intensity of the sound in
position $(i,j)$ at time $t$ is:
\begin{equation}
 \label{soundlevel}
S(i,j,t)=\sum_{i'=1}^n\sum_{j'=1}^n \delta(i',j',t) \cdot K(i-i')
\cdot K(j-j')
\end{equation}
where $K$ is the sound weakening factor (a 5-point derivative kernel
\cite{simoncelli_5point}), and $\delta(i',j',t')$ takes values 1 and
0 if at time $t$ the spectator at position $(i^\prime,j^\prime)$
claps or not her/his hands, respectively. We assume that at time $t$
the sound intensity in the $(i,j)$ position is $S(i,j,t)$ and
$\delta(i,j,t)=1$, meaning at that time the spectator at position
$(i,j)$ claps her/his hands. It can be shown that the $S(i,j,t)$
could give us information on the total number of synchronized
clapping spectators. At time $t'$ when $\delta(i,j,t')=0$ but
$S(i,j,t')>S(i,j,t)$ the sound heard at position $(i,j)$ is stronger
to that at time $t$ when $(i,j)$-spectator also acted. This emerges
the need of expressing her/his {\it Companionship}. The greater the
$S(i,j,t')/S(i,j,t)$ the higher the priority given to this action,
namely $C_{prio}$.

At any given moment $t$ a random number $S$ is generated: if
$S<(I_{prio}/(I_{prio}+C_{prio}))$  then the action (1) is chosen
else the action (2).

\begin{figure}
 \includegraphics[scale=0.5]{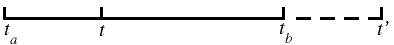}
 \caption{\label{fig2:tij}}
 \end{figure}

Let us see what happens when the action (2) is chosen. If $t$ is the
present time and $t_a$ the time (before $t$) when
$\delta(i,j,t_a)=1$ and $t_b$ the time (after $t$) when
$\delta(i,j,t_b)=1$ again, then $t_b =t_a + T_{i,j}$, with $T_{i,j}$
the self-period of the spectator $(i,j)$ (see Fig. \ref{fig2:tij}).


The following three cases might occur:
\begin{itemize}
\item if $t$ is closer to $t_a$, the spectator increases her/his self-period hoping that she/he
will be synchronized with the next clapping.
\item if $t$ is closer to $t_b$, the spectator lowers her/his self-period hoping that she/he will be synchronized with
the next clapping.
\item if $t$ is closer to $(t_a+t_b)/2$, the clapping at that time will result to an increment
of the background noise that is strongly related to the synchronization. To that extend, the $(i,j)$ wishes
not to increase the background noise thus suspends her/his clapping at time $t_b$ (we note that this
could happen only once).
\end{itemize}

\section{Experimental results - Discussion}
A statistical ensemble resembling a grid of 625 ($25 \times 25$)
spectators was simulated.  We begin the simulation with self-periods
following normal distribution with average value 50 and $\sigma=10$.
We assume that immediately after the clapping starts, spectators
feel a great need for expressing her/his enthusiasm or {\it
Individualism}, thus $I_{prior}$ is high. For this reason spectators
retain their self-periods of clapping and consequently there are no
signs of synchronization. When time passes $I_{prior}$ gets lower
resulting to the forming of random, at the beginning, clusters, that
give rise to the expression of {\it Companionship.} Immediately
after the forming of the first cluster an abrupt increase of the
average period is observed as it shown in Fig.
\ref{fig3:average_period}. This is in agreement with results
presented in
\cite{barabasi_nature_403_849_00,barabasi_pre_61_6987_00}.

The critical point $t_c$ of this abrupt change of period
($t_c\approx2000$) is characterized by a domination of the
$C_{prio}$ over the $I_{prio}$ and for this to happen, $I_{prio}$
has to be low enough, thus the incoherent clapping before
synchronization occurs. After some time, $I_{prio}$ starts to
increase because it reaches the $LST$. Spectators are starting to
express again themselves by clapping at their eigenfrequency
resulting to the loss of synchronization. Later, $I_{prio}$ starts
to decay because it reaches the $HST$ and synchronization might
occur again. This mechanism might be repeated several times during
the clapping period as it is found experimentally
\cite{barabasi_pre_61_6987_00}.

Figure \ref{fig4:sound_level} shows the sound level at the position
(12,12) of the audience. At the critical time $t_c$, synchronization
is visible as pronounced spikes in accordance to experimental
findings \cite{barabasi_pre_61_6987_00}. The dominant period after
time $t_c$ can be evaluated by counting the spikes resulting to a
value of $\approx 100$ which is in agreement with the findings of
Fig. \ref{fig3:average_period}.


Moreover, it seems that the mean period of the spectators is a
function of $LST$: the maximum mean period  recorded in the
simulation vs. the $LST$ is depicted in Fig.
\ref{fig5:space_period}. The error bars are proportional to the
background sound, which is a measure of synchronization as it is
explained in \cite{barabasi_nature_403_849_00}.


Finally, it is interesting to follow the dynamics of the abrupt
change of the mean period shown in Fig. \ref{fig3:average_period}.
Initially, lack of synchronization is due to the high value of the
$I_{prio}$ as it was explained before. After some time (the value of
$I_{prio}$ has been decreased), random clusters of synchronized
spectators may be formed producing locally a high sound level.
Spectators near those clusters either slowly start to take action to
synchronize with them or select to suspend some of their claps. In
the latter case the background noise is decreased (since some claps
are missing) causing $C_{prio}$ to increase suddenly. This effect
acts as an avalanche making the other spectators to act faster to
synchronize with the already formed clusters since their sound will
now be more evident. Hence, it is expected that the aforementioned
mechanism has a stronger effect on those spectators who clap at
higher frequencies, since these spectators contribute more to
background noise. For this reason, low clapping frequencies are
sustained and the dominant period during synchronization is
increased in comparison to the mean self period of the spectators.

Our proposal was further tested for its validity against the
Barab\'asi-Albert model \cite{
barabasi_science_286_509_99,barabasi_revmodphys_74_47_02} that
introduces the {\it ''preferential attachment''} concept in the
interpretation of the dynamics of large and complex networks. We
formulate the model by setting up the following differential
equation:
\begin{equation}
 \label{preferential}
\frac{\partial k_j}{\partial t} = a m \frac{1}{N-1} + (1-a) m
\frac{k_i}{\sum_j k_j}
\end{equation}
where $k_j, ~k_i$ is the degree of the respective node (the number
of edges starting from node $j$ or $i$), $t$ is the time step, $m$
is the number of links that $Nth$ node will establish, $N-1$ the
number of the existing nodes of the cluster, and $a$ is a parameter
that takes values as $0<a<1$; when $a=0(a=1)$ it is the
$Companionship(Individuality)$ that controls the direction of the
added links. In other words the left term from the right hand part
of Eq. \ref{preferential} counts for the expression of
$Individuality$ as: ''links will be placed randomly'', whereas the
right term counts for the expression of $Companionship$ as: ''links
will be directed to the node with the highest degree''. The above
Eq. \ref{preferential} can be produced by the following queuing
process; when a new node is added to the network, the m-edges that
carries are placed one by one as follows: The first node is placed
randomly with probability $p_1$ or following the {\it ''preferential
attachment'' } mechanism with probability $1-p_1$. Next, the second
edge is placed following the same scheme but with probability $p_2$,
and so on. It can be shown that $a=\frac{1}{m}\sum_{i=1}^mp_i$ and
Eq. \ref{preferential} holds. The solution of this differential
equation leads to the following:
\begin{equation}
 \label{preferential_solution}
 k_j(t) = \frac{1-\beta}{\beta} m \left ( \frac{t}{t_i} \right )^\beta - \frac{1-2\beta}{\beta} m
\end{equation}
where $\beta= \frac{1-a}{2}$ and $t_i$ the moment of insertion of the new node. Hence,
\begin{equation}
 \label{assumption}
 \frac{1-\beta}{\beta} m \left ( \frac{t}{t_i} \right )^\beta >> \frac{1-2\beta}{\beta} m
\end{equation}
 then the Eq. \ref{preferential_solution} is reduced to:
\begin{equation}
 \label{preferential_solution_final}
 k_j(t) = \frac{1-\beta}{\beta} m \left ( \frac{t}{t_i} \right )^\beta
\end{equation}
That is the distibution of degrees is given as:
\begin{equation}
 \label{distribution}
 P(t) \backsim 2 \left ( \frac{1-\beta}{\beta}\right )^{\frac{1}{\beta}} k^{\left (1+ \frac{1}{\beta}\right )}
\end{equation}
thus, the network is evolved by following a power law defined as
\begin{equation}
\gamma = 1 + \frac{1} {\beta} ~\text{or}~\gamma = 1+ \frac{2} {1-a}
> 3,~\text{for}~ a \neq 1
\end{equation}
In the extreme when $a=1$ the model reduces to random network
\cite{erdos_pubmath_6_290_59}, while for $a=0$ the model coincides
with the Barab\'asi-Albert model \cite{barabasi_science_286_509_99}.
The above described formalism enables us to produce {\it undirected
scale free networks} with $3<\gamma<\infty$. An illustrative example
comes from the {\it preferential attachment in sexual networks}. In
this particular case the value of $\gamma=3.4$ is the largest
observed
\cite{barabasi_revmodphys_74_47_02,liljeros_nature_411_907_01,liljeros_pnas_104_10762_07}
resulting to $a=16.7 \%$ or an expression of {\it Companionship}
which in this case is translated to:''the target sexual partner is
the one with the highest number of sexual links''. This argument
further supported by the fact that although preferential attachment
in sexual networks has been experimentally established
\cite{dugatkin_01}, the individual heterogeneity in the inclination
to find new partners was found essential to model experimental data
\cite{liljeros_pnas_104_10762_07}.

\section{Conclusions}
In summary, by augmenting the  V\'azquez et al.
\cite{barabasi_pre_73_036127_06} model with two contradicted {\it
driving forces} we were able to simulate the abrupt nature of
synchronization. In addition we have shown that since in the first
moments after the performance most spectators applause at their own
pace ($I_{prio}$ is large) there is no synchronization. Later, when
the value of $I_{prio}$ has been decreased, spectators may act to
express their $Companionship$. Random clusters of partial
synchronized spectators cause an abrupt increment of the $C_{prio}$
of the neighboring spectators, which in turns leads to ''global''
synchronization. However, after some ''synchronized'' period the
$I_{prio}$ could be enhanced again, resulting to lose of
synchronization. Last but not least the expression of {\it
Individualism} does not characterize communities of the Eastern
European countries where {\it Companionship} is the main driving
force, thus the easiness to achieve synchronization, whereas the
opposite is observed in Western countries communities
\cite{barabasi_nature_403_849_00}.

\bibliography{dx_dsv_tes_03_draft}
\clearpage
\section*{List of Figures}
\begin{figure}
 \includegraphics[scale=0.5]{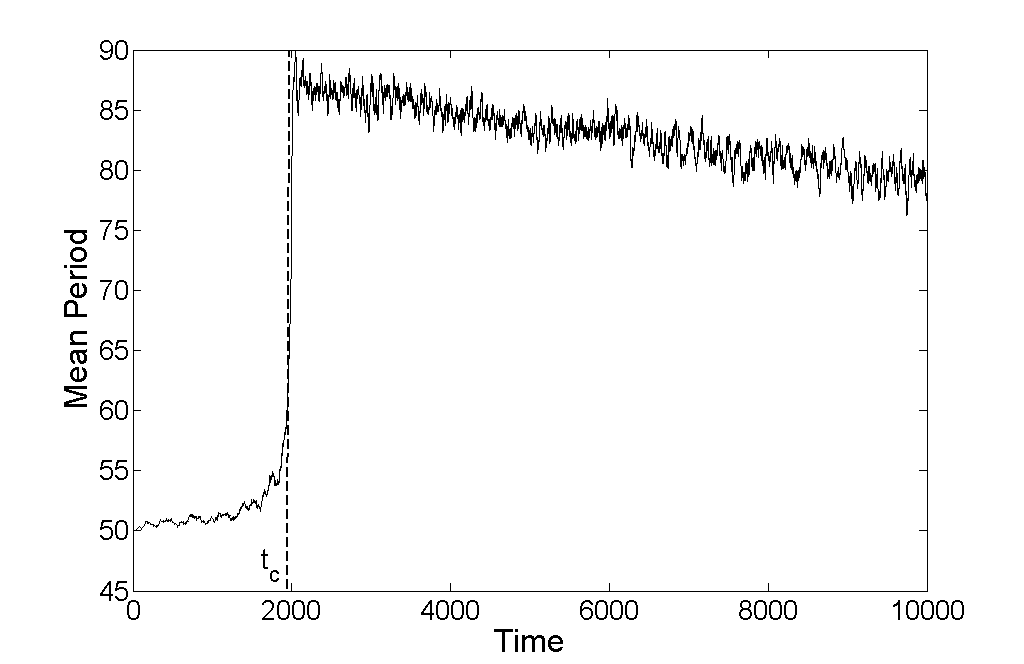}%
 \caption{\label{fig3:average_period}}
 \end{figure}

\begin{figure}
 \includegraphics[scale=0.5]{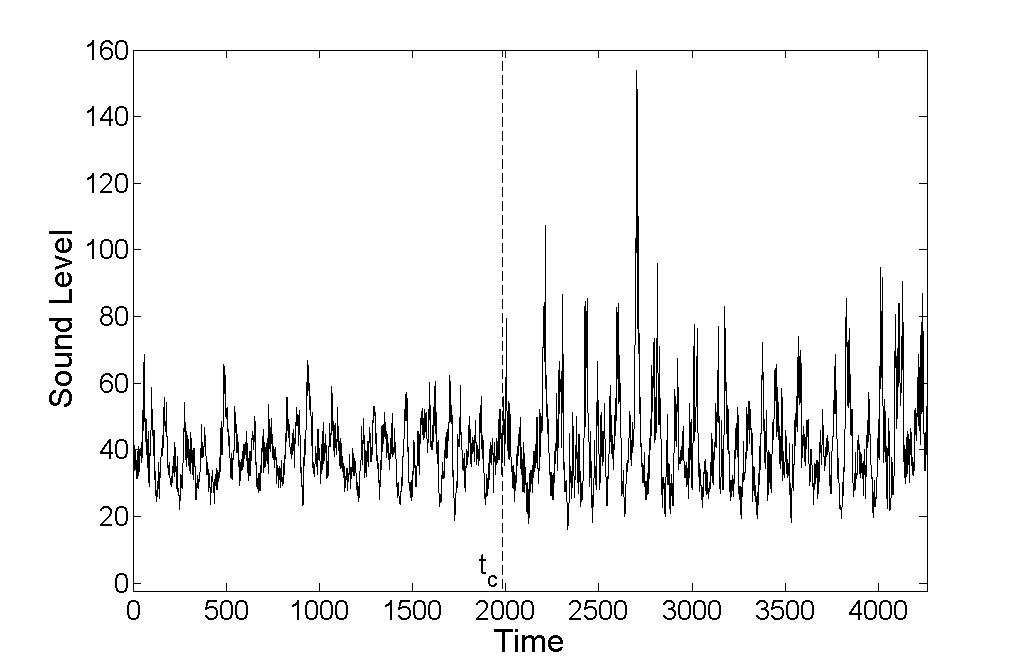}%
 \caption{\label{fig4:sound_level}}
 \end{figure}

\begin{figure}
 \includegraphics[scale=0.5]{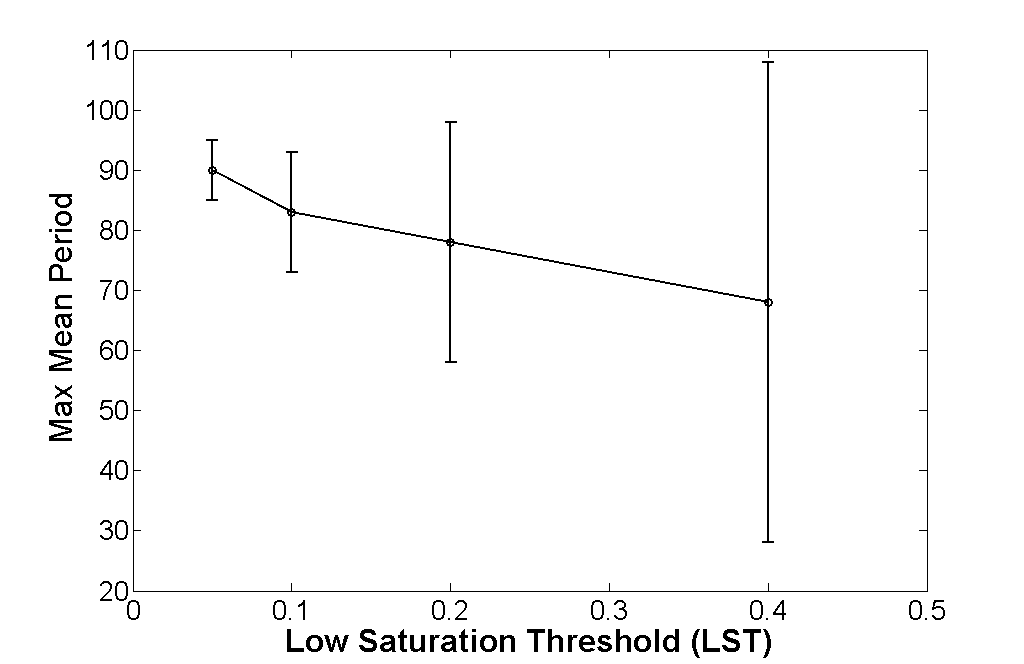}%
 \caption{\label{fig5:space_period}}
 \end{figure}

\begin{description}
\item [Fig. 1] The priority $I_{prio}$ of spectators to express their $Individualism$ as a function of time.
When $I_{prio}$ climbs over $HST$ then the need is considered
saturated and $I_{prio}$ starts to decay. The opposite behavior for
$I_{prio}$ is observed when it reaches $LST$, and thus the need
comes again to the foreground causing $I_{prio}$ to increase.

\item [Fig. 2] Different moments during the {\it decision making}  interval.

\item [Fig. 3] The average period of an audience of 625 spectators in a rectangular $25 \times 25$ grid.
The self-periods follow a normal distribution with mean value 50 and
$\sigma = 10$. The simulation parameters are: $HST=0.9$, $LST=0.05$,
$dI_{prio}/dt = \pm 0.001$.

\item [Fig. 4] The sound level at position (12,12) of the audience of the same simulation as in
Fig. \ref{fig3:average_period}. Synchronization after the critical time $t_c$ is indicated by the recorded spikes.

\item [Fig. 5] The maximum mean period recorded just after the critical time $t_c$, of the same simulation as
 in Fig. \ref{fig3:average_period}, vs. the level of the Low Saturation Threshold $LST$. The error bars are proportional
 to the background noise which is considered to be a measure of lack of synchronization.
\end{description}

 \clearpage
\clearpage
\clearpage
\clearpage
\clearpage
\end{document}